# Interaction between global-scale atmospheric vortices:

# Modeling with Hamiltonian dynamic system

# of antipodal point vortices on a rotating sphere

Igor I. Mokhov[1†], S. G. Chefranov[1]

and A. G. Chefranov[2]

[1]A.M. Obukhov Institute of Atmospheric Physics RAS, Moscow, Russia

[2] Eastern Mediterranean University, Famagusta, North Cyprus

## Abstract

We get point vortices dynamics equations on a rotating sphere surface directly from the hydrodynamic equations as representing their weak exact solution contrary to the conventional case of the use of a kinematic relationship between a given singular vortex field and velocity field. It is first time that the effect of a sphere rotation on the vortices interaction is accounted for exactly. We prove that only antipodal vortices (APV) pair with two point vortices in the diameter-conjugated points of a sphere with equal by quantity but different sign circulations, may be correctly considered in the frame of hydrodynamics equations as an elementary (stationary, not self-affecting) singular point object on a sphere. We suggest using the axis connecting the two point vortices in an APV for describing of an axis of rotation of the global vortices introduced in (Barrett, 1958) to reflect the observed global rotation of atmospheric masses with the rotation axes not coinciding with the planet rotation axis and precessing about it. Due to the non-linearity of the equations, there is no opportunity for consideration of the linear superposition of the Barrett's solutions (generalizing wave Rossby solutions). That is why, up to now, the question about interaction of global vortices corresponding to such solutions with rotations about different axis was not even posed. This difficulty, as we show, may be overcome if for the Barrett's vortices interaction modeling to use Hamiltonian equations of dynamics of N strongly interacting APV corresponding to an exact weak solution of hydrodynamics equations on a rotating sphere. More details are given to the case of N=2 when two Barrett's vortices are modelled. This is the first model describing interaction of the Barrett-type global vortices corresponding to atmospheric centers of action (ACA). The new steady-state and its stability conditions for $N$=2 are obtained and used for the analysis of coupled cyclone-anticyclone ACAs over oceans in the Northern Hemisphere. We prove that existence of such a stationary mode corresponding to the observed relatively static position of a cyclonic-anti-cyclonic global vortices pair above the ocean is a consequence of an exact accounting for the sphere rotation when considering point vortices dynamics. We show also an opportunity of existence of stable dynamically stationary modes for a pair of counter-sign vortices moving with the constant speed to the west. On the base of corresponding exact solutions, we show acceptability of modelling of the stable blocks of splitting flow type when in line with the exact accounting for the sphere rotation, we define also conditions of the polar vortices affecting on the stability boundaries of the block modes.

† E-mail address for correspondence: mokhov@ifaran.ru

1. **Introduction**

*1.1. Problem definition*

Investigation of the fluid dynamics on the rotating sphere has fundamental and applied significance for understanding of the key physical processes in the atmosphere and ocean. Studies on the base of the regular vortex waves and on the singular vortex models were conducted mostly independently. All of them must be based only on the solving of the same absolute vorticity conservation equation (AVCE) in a thin layer of ideal incompressible fluid on the surface of a rotating sphere.

The consideration of point vortices on a sphere, and the basics of this model were described by Gromeka (1952), Zermelo (1902) ), Bogomolov (1977, 1979), DiBatista & Polvani (1998), Dritschel et al. (2015) and others for the case of a static sphere and only for some approximate methods of the sphere rotation. The main conclusion in (Gromeka 1952) is that a dynamic system for the point vortices can be built only under necessary condition of zeroing the sum of intensities of all the point vortices. Under such a condition, there was suggested considering on the static sphere of " …a system of an even number of vortices having pairwise equal but inverse intensities and placed symmetrically with respect to some large circle". It corresponds to the system of pairs of antipodal point vortices (APV).

Up to now, equations describing point vortices dynamics on a sphere, were not inferred from the AVCE but were introduced using kinematic concepts on the base of the velocity field representation via definite singular vortex field distribution having a form of Bio-Savar-Laplace law. And up to now, there are no examples of exact accounting for the sphere rotation impact on the mutual dynamics of the point vortices.

Herein, this gap is filled. Directly for the AVCE, we obtain an exact weak solution in the form of Hamiltonian finite-dimensional dynamic system of ordinary differential equations. The system describes interaction of the point vortices under exact accounting for the effect of the sphere rotation. It is used for modelling of interaction of the global Barrett's vortices. Earlier, Barrett's vortices were considered only in the frame of study of regular wave solutions of the AVCE.

Emerging and developing of the approach based on the regular solutions of the AVCE on the rotating sphere was initiated by Rossby and is related to the study of Rossby vortex waves (Rossby 1939). We present in this paper the results which are related with those of Barrett (1958) that develop studies initiated by Rossby. The rest of the paper is organized as follows. In Section 1.2, we show that only APV pair satisfies the original three-dimensional hydrodynamics equations on a sphere. In Section 2, we present the AVCE and discuss its regular solutions in terms of Rossby vortex waves and the Barrett global vortices. In Section 3, we get an exact weak solution for the AVCE. This solution is presented as a solution of $2N$ Hamiltonian equations corresponding to $N$ APV pairs. Position of each APV on the sphere can characterize spherical coordinates of the Barrett's vortex axis. In Section 4, we consider a particular case of APV dynamics with $N=2$, stationary APV modes and their stability. In Section 5, we estimate variability (stability) of vortex modes from observations and from the model. Section 6 concludes the paper.

In this paper, we develop a new approach for modeling of the interaction of global-scale atmospheric vortices, some aspects of which were discussed earlier in (Mokhov et al. 2010, 2013). The obtained results can be used for diagnostics and modeling of atmospheric centers of action (ACAs) and blocks, in particular.

### 1.2. Problem of one point vortex on the sphere

When considering point vortices dynamics on a sphere, it is essential to find an elementary (stationary) singular vortex object character for the dynamics on the sphere surface. On a plane, such a vortex object is a point vortex without motion self-induction. Let us show that for the case of a spherical closed single-connected surface, a similar elementary vortex object can be only the dipole APV vortex, consisting of two point vortices of same by quantity but different sign intensity placed in the diameter-conjugated points of the sphere. The main conclusions of the present paper are based on the point vortices dynamics on a rotating sphere representation as a system of APVs.

Traditionally, as an elementary singular vortex object on the sphere, as on the unbounded plane, a unity point vortex is considered (Bogomolov 1977, 1979; Dritschel & Boatto 2015, and others); it has the following stream function (for simplicity, let a point vortex is placed on the sphere pole where the latitude complement to $\pi$, $\theta = 0$)

$$\psi_1 = \frac{\Gamma_1}{4\pi} \ln(\frac{1}{1-\cos\theta}). \tag{1.1}$$

Instead of (1.1) in this paper we use stream function for APV-in-the next form:

$$\psi = \frac{\Gamma_1}{2\pi} \ln(\frac{1+\cos\theta}{1-\cos\theta}). \tag{1.2}$$

Let us show that contrary to the stream function (1.1), the stream function (1.2) satisfies the original tree-dimensional hydrodynamics equations (used for inference of AVCE in a thin layer of ideal incompressible fluid on a rotating sphere, see below equation (2.1)).

In the case of a stationary mode when only zonal velocity field component is non-zero, $V_\varphi = -\frac{1}{r}\frac{\partial \psi}{\partial \theta}$, hydrodynamics equations of ideal incompressible fluid in spherical coordinates, $(r, \theta, \varphi)$, is as follows (here, for simplicity, we consider the case of a static sphere in axisymmetric case when dependence on angle, $\varphi$, is absent)

$$\frac{V_\varphi^2}{r} = \frac{1}{\rho_0}\frac{\partial p}{\partial r}, \tag{1.3}$$

$$\frac{V_\varphi^2}{r} ctg\theta = \frac{1}{r\rho_0}\frac{\partial p}{\partial \theta}. \tag{1.4}$$

The condition of compliance of equations (1.3) and (1.4) is as follows (it is consequence of the equality $\frac{\partial^2 p}{\partial \theta \partial r} = \frac{\partial^2 p}{\partial r \partial \theta}$ and is obtained by differentiation of the both sides of the equation (1.3) over $\theta$ and differentiation of the both sides of the equation (1.4) over r)

$$\frac{1}{r}\frac{\partial V_\varphi}{\partial \theta} = \frac{\partial V_\varphi}{\partial r} ctg\theta . \qquad (1.5)$$

For the APV stream function (1.2), we have $V_\varphi = \frac{\Gamma_1}{r\pi \sin\theta}$, that, as not difficult checking, satisfies (1.5) for any $r; \theta$. At the same time, for the stream function (1.1), we have

$$V_\varphi = \frac{\Gamma_1}{4r\pi} ctg(\frac{\theta}{2}) . \qquad (1.6)$$

Substituting (1.6) in (1.5), we get that the velocity field (1.6), corresponding to the stream function (1.1), does not satisfy condition (1.5) for any $\theta$ and for that velocity field hydrodynamics equations (1.3) and (1.4) are not compatible.

Thus, the stream function (1.1) cannot describe any realizable hydrodynamic flow contrary to the APV stream function of the form (1.2).

## 2. Rossby waves and Barrett vortices

*2.1. Absolute vorticity conservation equation*

Rossby (1939) selected long-living quasi-stationary planetary-scale structures, ACAs, after filtering out pressure field fluctuations related with mobile cyclones. To simulate ACA type modes, linear solution of absolute vorticity ($\omega$) equation was used by Rossby in the $\beta$-plane approximation. Further consideration of this problem was conducted by Haurwitz (1940), Craig (1945), Neamtan (1946) and Barrett (1958) on a rotating sphere of radius $R$. The equation of $\omega$ conservation, AVCE, in spherical coordinates ($r$, $\theta$, $\phi$) is in this case as follows (Batchelor 1967) (for $r=R$):

$$\frac{D\omega}{Dt} = \frac{\partial \omega}{\partial t} + \frac{V_\theta}{R}\frac{\partial \omega}{\partial \theta} + \frac{V_\varphi}{R\sin\theta}\frac{\partial \omega}{\partial \varphi} = 0. \tag{2.1}$$

Here: $\omega = \omega_r + 2\Omega\cos\theta$, $\Omega$ is the angular velocity of the sphere rotation (for the Earth $\Omega \approx 7.3 \cdot 10^{-5} \sec^{-1}$), $\theta$ is the co-latitude, $\varphi$ is the longitude; $V_\theta = R\dot\theta$, $\dot\theta = \frac{d\theta}{dt}$, $V_\varphi = R\sin\theta\dot\varphi$, $\omega_r = \frac{1}{R\sin\theta}(\frac{\partial(V_\varphi \sin\theta)}{\partial \theta} - \frac{\partial V_\theta}{\partial \varphi}) = -\Delta\psi$ is the radial component of the local vortex field on the sphere, $\Delta$ is Laplace operator; $\psi$ is the stream function for which $V_\varphi = -\frac{1}{R}\frac{\partial \psi}{\partial \theta}$, $V_\theta = \frac{1}{R\sin\theta}\frac{\partial \psi}{\partial \varphi}$ in (2.1). Equation (2.1) in the more general case corresponds to potential vorticity conservation $\frac{D}{Dt}(\frac{\omega}{H}) = 0$, where $H$ is the thickness of the fluid layer (see Batchelor (1967)). Equation (2.1) holds not only for the constant $H$, but also when $H$ is a Lagrangian invariant and the velocity field (with zero radial velocity component $\dot r = V_r = 0$) is non-divergent i.e. $divV = \frac{1}{R\sin\theta}(\frac{\partial V_\theta \sin\theta}{\partial \theta} + \frac{\partial V_\varphi}{\partial \varphi}) = -\frac{1}{H}\frac{DH}{Dt} = 0$. The latter allows to introduce the stream function $\Psi$. Note that relative (local) vorticity $\omega_r$ varies only for the fluid element moving with the change of latitude. Equation (2.1) is applicable to flows with any character length scale $L$ only when $L > H$ for $H \ll R$.

**2.2. Solutions of AVCE (2.1) by Rossby (1939), Craig (1945) and Neamtan (1946)**

The solution obtained for equation (2.1) by Barrett (1958) (as well as by Rossby (1939)), in a linear approximation has the following form

$$\psi = \alpha R^2 \cos\theta + \psi_{\alpha 0}\cos(\beta t + m\varphi)P_n^m(\cos\theta), \tag{2.2}$$

$$\beta = m(\frac{2(\alpha + \Omega)}{n(n+1)} - \alpha) \qquad (2.3)$$

with integer numbers $n$ and $m$, $m \leq n$, $n = 1, 2, ..$, and adjoint Legendre polynomials $P_n^m$. It was assumed by Rossby (1939) that the disturbance amplitude $\psi_{\alpha 0}$ is small with respect to the value $\alpha R^2$, characterizing the zonal flow intensity $V_{0\varphi} = \alpha R \sin\theta$, relative to which wave disturbance for the vortex field is considered based on equation (2.1).

In (Craig 1945) and (Neamtan 1946), it is shown that solutions of the type (2.2), (2.3) also preserve their form for the general case of nonlinear waves when the condition $\Psi_{\alpha 0} << \alpha R^2$ is not necessary. The solutions represented by equations (2.2), (2.3) in the limit $\alpha \to 0$ describe a wave traveling from east to west with an angular velocity $c = \frac{d\varphi}{dt} = -\frac{2\Omega}{n(n+1)}$. Steady-state condition with $\beta = 0$ in (2.3) for such a wave are necessary for the description of ACA type modes that are found to be possible only for a definite value of $\alpha = \frac{2\Omega}{n(n+1) - 2}, n > 1$ and corresponding to very special zonal flow intensity (see first member in (2.2)). The latter is usually related to the meridional temperature gradient. Thus the ACA description on the basis of equations (2.2), (2.3) is possible only with the additional determination of the value for the free parameter $\alpha$, which is not defined in the hydrodynamic model of equation (2.1). As a result, a pure hydrodynamic description of the quasi-stationary structures of ACA type is not possible on the basis of the solution from equation (2.1) in the form of equations (2.2), (2.3). The solution for this problem is proposed in Section 3.

### 2.3. Solutions of AVCE(2.1) by Barrett (1958)

In (Barrett 1958), a new solution form of the nonlinear equation (2.1) is obtained. This solution has corresponding vortex motion of the global scale embracing entire Earth atmosphere. The axis of such a global vortex position of which is specified by spherical coordinates $(\theta_0, \varphi_0)$ at the initial time instance, does not coincide with the sphere rotation axis and .has a precession about the axis of the sphere rotation from east to west with an angular velocity of $c = -\dfrac{2\Omega}{n(n+1)}$, where $\Omega$ is the Earth rotation frequency. Given that in (Barrett 1958) the form the solution takes (after substitution of $\cos u_0 = \cos\theta\cos\theta_0 + \sin\theta\sin\theta_0\cos(\varphi - \varphi_0)$ instead of $\cos\theta$ in equations (2.2), (2.3) for $\alpha = 0$ in (2.2), (2.3)) the following form results:

$$\psi = A_1 + B_1[P_n(\cos\theta_0)P_n(\cos\theta) + 2\sum_{m=1}^{n}\frac{(n-m)!}{(n+m)!}P_n^m(\cos\theta_0)P_n^m(\cos\theta)\cos m(\varphi - \varphi_0 + \frac{2\Omega t}{n(n+1)}) = \quad (2.4)$$
$$= A_1 + B_1 P_n(\cos u_0).$$

Here $\theta_0, \varphi_0$ are spherical coordinates of the initial axis position for the Barrett-type planetary-scale vortex, introduced by Barrett (1958) in correspondence to the observations from La Seur (1954). Solution (2.4) was suggested by Barrett (1958) for description of the observed global rotations of atmospheric masses with rotation axes not coinciding with the planet rotation axis and precessing about it in the west direction. According to La Seur (1954), the circulation of the middle troposphere possesses a remarkable eccentricity with the center of symmetry for the flow at a considerable distance from the geographical pole. In such a case the intensity of the zonal component of the flow (with respect to the geographic pole) is low, while the meridional component is large. Harmonic analysis shows that the most of the energy for meridional motion is associated with the first longitudinal harmonic (Barrett 1958). According to Barrett (1958) the circulation with a large meridional component may be a rather symmetric zonal vortex with respect to an eccentric pole.

As noted in (Barrett 1958), such eccentric planetary-scale vortices were not considered previously in contrast with vortices (with significantly smaller scales) as introduced by Rossby (1948, 1949). Only for $N = 1$ does Barrett's planetary-scale vortex axis (which is static in the absolute coordinate system, where sphere rotates from west to east with angular velocity $\Omega$) rotate from east to west with angular velocity $c = -\Omega$. This case (with $N=1$ as in Barrett's consideration) is important for a better understanding of the new results of Section 2.5 and Section 3 of this paper. In Section 3, we consider *N* antipodal vortex pairs for modeling *N* axis positions and dynamics of *N* different Barrett-type planetary-scale vortices in a process with strong vortices interaction. Barrett (1958) considered only the case of one planetary vortex (*N*=1). The first model consideration of interaction between any number of Barrett-type planetary-scale vortices and of more than two (*N*>2) is presented in Section 3.

**2.4. Solutions of AVCE (2.1) by Verkley (1984)**

For the further consideration it is convenient to represent the solution of the equation (2.1) as obtained by Barrett (1958), but in the following more generalized form, allowing singularities in the structure of the corresponding to such a solution (2.1) vortex wave embracing entire atmosphere (see also (Verkley 1984)):

$$\psi = Y\left(\cos u_0\left(\theta,\ \theta_0,\ \varphi-\varphi_0+\frac{2\Omega t}{\nu(\nu+1)}\right)\right). \quad (2.5)$$

In (2.5), *Y* is the eigen-function of the Laplace operator, i.e. $\Delta Y = -\frac{\nu(\nu+1)}{R^2}Y$. The case of integer eigen values $\nu = n$ corresponds to the solution (2.4). For arbitrary (including complex) $\nu$, a solution of similar type was analyzed by Verkley (1984) in relation to the modeling of blocking events based on smaller-scale dipole vortex structures (modons) constructed with the use of functions $Y$. In Verkley (1984), the case was investigated by defining $Y$ in equation (2.5) in terms of Legendre functions of the first and second kind (the second kind function $Q_0$, which have singular behavior at two points on sphere, we shall also consider in Section 3): $P_\nu^\gamma, Q_\nu^\gamma$, i.e. $Y(\theta,\varphi) = G(\varphi)H(\theta), G = e^{\pm i\gamma\varphi}, \gamma = m, m = 0, \pm 1,..,\ H(\theta) = \begin{cases} P_\nu^\gamma(\cos\theta) \\ Q_\nu^\gamma(\cos\theta) \end{cases}.$

It is worth noting that for the stationary case with $\Omega = 0$ (i.e. for static sphere) there exists similar to equation (2.5), but an even more general form (as suggested by E.A. Novikov, personal communication) for the solution of (2.1) as $\psi = F(\cos u_0)$, where $F$ is an arbitrary function of $\cos u_0(\theta, \theta_0, \varphi - \varphi_0)$ (see (Chefranov 1985)). In particular, in (Chefranov 1985) a model for global pollution transport in a statistical ensemble of the ACA-type vortices is considered as an example of a linear function $F$ corresponding to solid-body rotation.

## 2.5. Generalized Barrett (1958) and Verkley (1984) solutions of AVCE (2.1)

It is also possible to get a new modification of solution (2.5) in which the precession frequency of the Barrett vortex axis may be already independent from the eigenvalue $\nu$. Let us represent the solution of (2.1) as a linear superposition (in the most general form when $\alpha \neq 0$) of the stream function $\psi_0$ and (2.5):

$$\psi = \psi_0 + Y + \alpha R^2 \cos\theta. \qquad (2.6)$$

The function $\psi_0$ in equation (2.6) corresponds to zero absolute vorticity $\omega = -\Delta\psi_0 + 2\Omega\cos\theta = 0$ and has a form $\psi_0 = -\Omega R^2 \cos\theta$ characterizing solid-body east to west fluid rotation relative to the sphere surface. In this case the fluid in the absolute coordinate system (in which the sphere is rotate with angular velocity $\Omega$) is static for $\omega = 0$ and $\psi = \psi_0$.

For the stream function, in form (2.6) with any $\nu$ in $Y$, the corresponding global vortices' axes rotate with the same angular velocity $\Omega$ from east to west only in the case $\alpha = 0$. In contrast to the stream function in form (2.5), this velocity does not depend on the value $\nu$ (or $n$). Actually, from equations (2.1) and (2.6), it follows $\frac{\partial Y}{\partial t} + c\frac{\partial Y}{\partial \varphi} = 0$ for $c = \alpha\left(1 - \frac{2}{\nu(\nu+1)}\right) - \Omega$. When $\psi_0$ is not introduced as in (Neamtan 1946) and (Verkley 1984), we have $c = \alpha\left(1 - \frac{2}{\nu(\nu+1)}\right) - \frac{2\Omega}{\nu(\nu+1)}$, where $\nu = n$ in (Neamtan 1946). Let us note that similar expression $c = \alpha\left(1 - \frac{2}{\nu(\nu+1)}\right)$ was obtained in (Zermelo 1902) for $\Omega = 0$ and $\nu = n$. In (Craig 1945) and ((Barrett 1958) for $\alpha = 0$ and $\nu = n$ the expression for angular velocity has a form $c = -\dfrac{2\Omega}{\nu(\nu+1)}$. It means that for the stream function in form (2.6) in absolute coordinate system Barrett planetary vortex axis (Barrett 1958) may be static for $\alpha = 0$ because

$$c = -\Omega \qquad (2.7)$$

when $\psi_0 \neq 0$ in (2.6).

In (2.7), already Barrett vortex axis precession angular velocity dependence on the parameter $\nu$ is absent. Consequently, for $\psi_0 \neq 0$ and $\alpha = 0$ in (2.6), all symmetry types Barrett vortices axes (specified by different parameter $\nu$ values) precess about the planet rotation axis with the same angular velocity equal by value but having an opposite sign to that of the planet. It means that in the absolute reference frame positions of axes of such vortices are static in the absence on inter-vortex interactions. We get an exact weak solution of the equation (2.1), allowing modelling of interactions between multiple Barrett vortices positions of axes of which are defined by respective coordinates of multiple APV on the sphere described by a Hamiltonian dynamic system.

Further in Section 3 the modeling of interaction of ACA type structures corresponding to the Barrett-type planetary vortices is proposed. We use a new approach for description of the point vortices on a rotating sphere based on the model corresponding to the equation of hydrodynamics on the sphere (2.1).

**3. Exact weak solution for AVCE (2.1)**

1. We shall model the Barrett vortices discussed in the previous section each as a pair of equal but opposite sign intensity antipodal singular vortices located on the sphere in the diameter-conjugated points. The dynamic interaction of such vortex pairs will be considered on the basis of an exact weak solution of the AVCE (2.1) on a rotating sphere.

It is possible to search a weak solution for equation (2.1) in the form of a superposition of $N+1$ diametrically conjugated point vortex pairs on the rotating sphere, APV (Borisov et al. 2007):

$$\omega = \frac{\Gamma_0}{R^2}(\delta(\theta)-\delta(\theta-\pi)) + \sum_{i=1}^{N}\frac{\Gamma_i}{R^2 \sin\theta_i}(\delta(\theta-\theta_i)\delta(\varphi-\varphi_i)-\delta(\theta-\pi+\theta_i)\delta(\varphi-\varphi_i-\pi)), \quad (3.1)$$

where $\delta$ is the Dirac delta-function. The stream function $\psi$ for the vorticity represented by equation (3.1) has the form of type (2.6) and corresponds to the presence of singularities in $2(N+1)$ points of the sphere:

$$\psi = \psi_0 + \frac{\Gamma_0}{\pi}Q_0(\cos\theta) + \sum_{i=1}^{N}\frac{\Gamma_i}{\pi}Q_0(\cos u_i(\theta,\varphi)), \quad (3.2)$$

where $\psi_0 = -\Omega R^2 \cos\theta$ (as in (2.6), introduction of that function corresponding to zero solution of the equation (2.1) allows exactly accounting for the sphere rotation), and $Q_0(x) = \frac{1}{2}\ln\frac{1+x}{1-x}$ is the Legendre function of the second kind of zero order and zero power, $\cos u_i = \cos\theta\cos\theta_i + \sin\theta\sin\theta_i\cos(\varphi-\varphi_i)$, and $\theta_i, \varphi_i$ are the spherical coordinates of the point vortices which may depend on time. Let us note that the pointed form of the Legendre function of the second kind in (3.2) corresponds to the particular case of introduced in (2.5) and (2.6) generalization of a Barrett vortex description. The use in the present paper of the very APV system is caused mainly by consideration of a problem of modelling of Barrett vortices interaction characterized by a global rotation axis position of which is adequately described exactly by an APV structure.

Obviously, the structure of expression (3.2) exactly corresponds to the particular case of the stream function (2.6) for $\alpha = 0$ and $m = n = 0$. The equation for the required zero integral vorticity on the sphere ($\int_0^{2\pi} d\varphi \int_0^\pi d\theta \sin\theta \omega = 0$) holds identically for any values $\Gamma_0$ and $\Gamma_i$, $i = 1...N$, in the representation form (3.1) for weak solution of equation (2.1).

The expression for $\psi$ in (3.2) takes into account (see also (1.2)) that solution of the equation

$$\omega_{r0} = -\frac{1}{R^2 \sin\theta}\frac{\partial}{\partial\theta}\sin\theta\frac{\partial\psi_{0V}}{\partial\theta} = \frac{\Gamma_0}{R^2}\left(\delta(\theta) - \delta(\theta - \pi)\right) \qquad (3.3)$$

is the function $\psi_{0V} = \frac{\Gamma_0}{2\pi}\ln\frac{1+\cos\theta}{1-\cos\theta}$. The last term in (3.2) is obtained by solving equation (3.3) using symmetry considerations with the substitution of $\cos\theta \to \cos u_i(\theta,\varphi)$ when the vortex axis turns in the direction of $(\theta_i, \varphi_i)$ from $\theta = 0$.

The vorticity field (3.1) and stream function (3.2) may be used for obtaining an exact weak solution (in the sense of generalized functions) for equation (2.1). As a result, the form is defined for functions $\theta_i(t)$, $\varphi_i(t)$ which are solutions of the following *2N*-dimensional Hamiltonian system of ordinary differential equations ($\dot\theta_i = \frac{d\theta_i}{dt}$ and so on):

$$\dot\theta_i = \frac{1}{R^2\sin\theta_i}\frac{\partial\psi(\theta_i,\varphi_i)}{\partial\varphi_i} = -\frac{1}{\pi R^2}\sum_{k=1,k\neq i}^{N}\frac{\Gamma_k \sin\theta_k \sin(\varphi_i - \varphi_k)}{1 - \cos^2 u_{ik}}, \qquad (3.4)$$

$$\sin\theta_i \dot\varphi_i = -\frac{1}{R^2}\frac{\partial\psi(\theta_i,\varphi_i)}{\partial\theta_i} = -\Omega\sin\theta_i + \frac{\Gamma_0}{\pi R^2 \sin\theta_i} -$$
$$\frac{1}{\pi R^2}\sum_{k=1,k\neq i}^{N}\frac{\Gamma_k(\cos\theta_i \sin\theta_k \cos(\varphi_i - \varphi_k) - \sin\theta_i \cos\theta_k)}{1 - \cos^2 u_{ik}}, i = \overline{1,N}$$

(see Appendix A.1 for more information).

Here $\Gamma_k = const$ for any $\Omega$=const and $\Gamma_0$=const (in principle both $\Omega$ and $\Gamma_0$ may depend on time), and $\cos u_{ik} = \cos\theta_i \cos\theta_k + \sin\theta_i \sin\theta_k \cos(\varphi_i - \varphi_k)$.

In our approach, point vortices dynamics is fully defined by the system (3.4) solution and does not depend on the fluid layer height $H$. Vice versa, the function $H$ must satisfy the following equation (see also the end of the Section 2.1):

$$\frac{DH}{Dt} = \frac{\partial H}{\partial t} + \frac{V_\theta}{R}\frac{\partial H}{\partial \theta} + \frac{V_\varphi}{R\sin\theta}\frac{\partial H}{\partial \varphi} = 0,$$

where the velocity field components $V_\theta, V_\varphi$ are determined by the stream function (3.2) which takes into consideration the solutions of the system (3.4) for the coordinates of all point vortices on the rotating sphere. Accounting for the sphere rotation is made in (3.2) due to the introducing in (3.2) of the stream function $\psi_0$.

More general representation of $\psi_0$ in the form of the new function $\psi_{0H}$ (defined from the equation (A.5) solution, see Appendix A.2) can be obtained on the base of consideration of potential vorticity conservation equation (A.4). Dynamics of the point vortices is defined from the generalizationof the system (3.4) and is already mutually closely related with variability of the layer height (defined from the joint solution of the equations(A.8) and (A.5) and taking into account the stream function (A.9)).

The system (3.4) for $\Omega = 0$ and $\Gamma_0 = 0$ with accuracy up to numerical factor (set at $\pi$) coincides with the corresponding system derived for the system of N pairs of APV in (Borisov et al. 2007), where the system is introduced using kinematic considerations (see (Zermelo 1902; Bogomolov 1977)), but not on the basis of an exact weak solution of hydrodynamics equation (2.1).In the case under consideration, the dissipation processes and energy pumping within the system are treated as not significant or balancing each other. Equations (3.4), following from (2.1) and (3.1), (3.2), must provide conservation of integral invariants of kinetic energy $\overline{E}$, angular momentum $\overline{M}$ and impulse $\overline{P}$, where the line above the variables denotes the averaging process for the respective variable on the surface of the sphere. Noted values related to the mass unit in the rotating coordinate system have the following form: $E = \frac{1}{2}V^2$, $\mathbf{M} = [\mathbf{r} \times \mathbf{V}]$, $\mathbf{P} = \mathbf{V}$, where $\mathbf{V} = \frac{d\mathbf{r}}{dt}$, $\mathbf{r}$ is the radius-vector in Cartesian coordinate system (x,y,z), origin of which is in the center of the sphere. The rotation of the sphere with a frequency of $\Omega$ is performed about the z-axis and radial motion is absent ($V_r=0$).

From definition $\bar{E} = \int_0^{2\pi} d\varphi \int_0^{\pi} d\theta \sin\theta E = \frac{1}{2}\int_0^{2\pi} d\varphi \int_0^{\pi} d\theta \sin\theta \psi \omega_r$ it follows that $\bar{\mathbf{P}} = 0$, and also:

$$\bar{E} = \frac{1}{8\pi}\sum_{i=1}^{N}\sum_{\substack{k=1,\\i\neq k}}^{N}\frac{\Gamma_i\Gamma_k}{R^2}\ln\frac{1+\cos u_{ik}}{1-\cos u_{ik}} + \frac{\Gamma_0}{2\pi R^2}\sum_{k=1}^{N}\Gamma_k \ln\frac{1+\cos\theta_k}{1-\cos\theta_k} + \frac{4\pi}{3}\Omega^2 R^2 - \Omega(2\Gamma_0 + \sum_{i=1}^{N}\Gamma_i\cos\theta_i), \quad (3.5)$$

$$\bar{M}_z = 2\sum_{i=1}^{N}\Gamma_i\cos\theta_i + 4\Gamma_0 - \frac{8\pi\Omega R^2}{3}, \quad (3.6)$$

$$\bar{M}_x = 2\sum_{i=1}^{N}\Gamma_i\cos\varphi_i\sin\theta_i, \quad \bar{M}_y = 2\sum_{i=1}^{N}\Gamma_i\sin\varphi_i\sin\theta_i. \quad (3.7)$$

Values of $\theta_i$ and $\varphi_i, i = \overline{1,N}$, in (3.5)-(3.7) are functions of time derived from the dynamic equations (3.4) under corresponding initial conditions. For $\Omega = 0$ and $\Gamma_0 = 0$ all four values (3.5)-(3.7) are invariants of the system (3.4). It may be checked by differentiating (3.5)-(3.7) with respect to time and taking into account (3.4). The invariants for the case $\Omega = 0, \Gamma_0 = 0$ exactly coincide with the invariants of the dynamic system in (Bogomolov 1977; Bogomolov 1979; Borisov et al. 2007).

For $\Omega \neq 0$ or $\Gamma_0 \neq 0$ values (3.7) are already not invariant, but values $\bar{E}$ and $\bar{M}_z$ are still invariants of (3.4) (it is true for $\bar{E}$ only if $\Gamma_0 = const$ and $\Omega = const$). Thus, the sphere rotation or accounting for the polar vortices removes degeneration corresponding to the symmetry-related condition with two invariants of (3.7), existing only for $\Omega = 0$ and $\Gamma_0 = 0$. As a result, the system (3.4) has for $\Omega \neq 0$ or $\Gamma_0 \neq 0$ only two independent integral invariants. Moreover, and for $\Omega = \Gamma_0 = 0$ the invariants of (3.7) may be not independent from (3.5), (3.6).

## 4. Stationary vortex modes ($N=2$) and their stability

Let us consider conditions of the existence and stability of the stationary modes corresponding to equilibrium for the vortex pairs for $N=2$ in (3.4). Under the condition $\dot{\theta}_1 = \dot{\theta}_2 = 0$ the equilibrium is possible either for $\varphi_1 = \varphi_2 = const$, or for $\varphi_1 - \varphi_2 = \pi$. It corresponds generally to two distinct vortex stationary modes on a sphere. For example, in the case $\varphi_1 = \varphi_2$, when $\dfrac{d(\varphi_1 - \varphi_2)}{dt} = 0$, the following condition is obtained:

$$\frac{\gamma_0(\sin^2\theta_{20} - \sin^2\theta_{10})}{\sin\theta_{10}\sin\theta_{20}} = \frac{\gamma_1\sin\theta_{10} + \sin\theta_{20}}{\sin(\theta_{20} - \theta_{10})}, \qquad (4.1)$$

where $\gamma_0 = \dfrac{\Gamma_0}{\Gamma_2}; \gamma_1 = \dfrac{\Gamma_1}{\Gamma_2}$ and $\theta_{10}$, $\theta_{20}$ are the stationary values of $\theta_1$ and $\theta_2$. An additional condition for the absence of absolute motion in this case using the equality $\dfrac{d(\varphi_1 + \varphi_2)}{dt} = 0$, has the form

$$\frac{2\pi R^2 \Omega \sin\theta_{10}\sin\theta_{20}}{\Gamma_2} = \frac{\gamma_0(\sin^2\theta_{10} + \sin^2\theta_{20})}{\sin\theta_{10}\sin\theta_{20}} + \frac{\gamma_1\sin\theta_{10} - \sin\theta_{20}}{\sin(\theta_{20} - \theta_{10})}. \qquad (4.2)$$

It is possible to show that for $\gamma_0 = 0$ a stationary vortex mode from (4.1), (4.2) existing for $\gamma_1 = -\dfrac{\sin\theta_{20}}{\sin\theta_{10}} < 0$, $\dfrac{\Omega R^2}{\Gamma_2} < 0$, is stable with respect to small disturbances. Note that the left hand side of (4.1) may turn to zero also for $\gamma_0 \neq 0$, when the relationship (3.13) holds. Respective stationary mode and condition (3.17) of its linear stability were discussed above in relation with the problem block events modelling.

For non-zero polar vortices intensity, their impact on stability of the stationary mode (4.1), (4.2) may be very substantial. On Fig. 1, we show small disturbances of the stationary mode for different values of the parameter, $g_0 = \dfrac{\Gamma_0}{\pi R^2 \Omega}$, characterizing the polar intensity. From Fig. 1, it is seen that when increasing the polar vortices intensity, the stable mode of nearly stationary position of a vortex pair modelling a block with split flow, may change to an unstable sufficiently fast pair motion in the west direction.

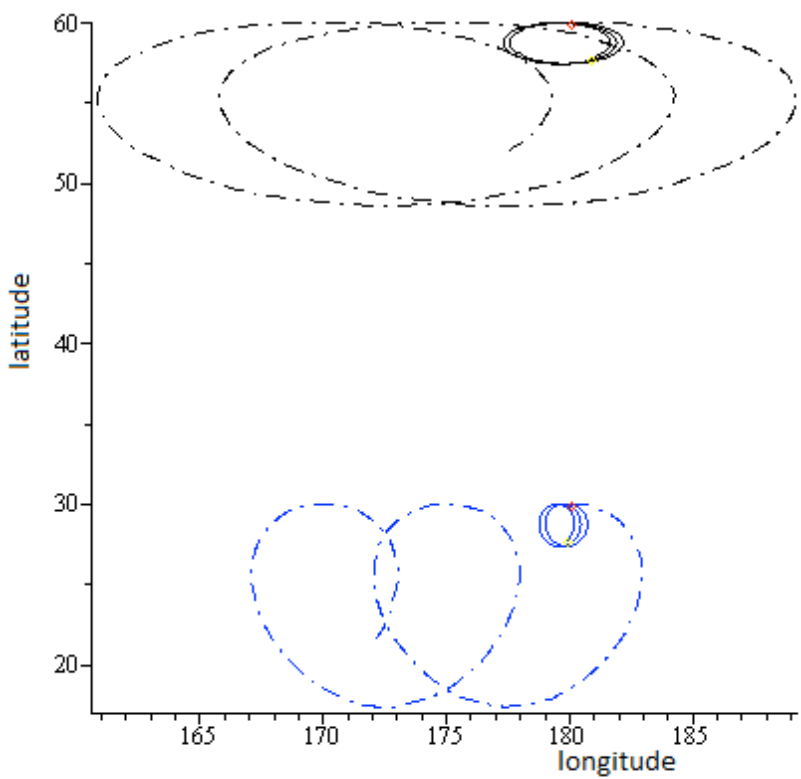

Fig. 1. Illustration of the system (3.4) solution for $N=2$. Dynamics of two APV on a sphere for 15 days for the stable mode (with small polar vortices intensity, $g_0 = 0.01$), solid lines, and unstable mode (with five times greater polar vortices intensity, $g_0 = 0.05$), point-dash lines. Initial vortices' coordinates in the form (longitude, latitude), in degrees, are (180, 60) and (180, 30). For the dimensionless cyclonic vortex intensity (it is located to the north), $g_1 = \sqrt{3}/4$, and for the anti-cyclonic vortex, $g_2 = -0.25$.

Let us consider a problem of stability of the stationary mode (4.1), (4.2) in the general case for $\gamma_0 = \dfrac{g_0}{g_2} = \dfrac{\Gamma_0}{\Gamma_2} \neq 0$.

It is possible to show that for $\gamma_0 \neq 0$ the considered stationary state is stable with respect to small disturbances only when the following inequality holds:

$$D \equiv A\gamma_1^2 + 2B\gamma_1 + C < 0, \qquad (4.3)$$

where

$$\theta_{10} \equiv y, \theta_{20} \equiv z, \ A = \sin^3 y [\sin(2z - y) + \frac{2\sin^2 y \cos z \sin(z - y)}{\sin^2 z - \sin^2 y}], \ B = \sin y \sin z (\sin^2 z + \sin^2 y)$$

$$C = \sin^3 z [\sin(2y - z) + \frac{2\sin^2 z \cos y \sin(z - y)}{\sin^2 z - \sin^2 y}].$$

Inequality (4.3) corresponds to the realization of an oscillatory mode for small disturbances described by the following system of equations:

$$\frac{d\theta_1^*}{d\tau} = -\Delta\varphi \sin z / \sin^2(z - y), \ \frac{d\Delta\varphi}{d\tau} = -\frac{\theta_1^* D}{\sin^2(z - y)\sin^2 y \sin^3 z},$$

where $\theta_1^*(\tau)$ is the disturbance of the stationary state $y$, and $\Delta\varphi(\tau)$ is the deviation from 0 for the longitude difference $\varphi_1 - \varphi_2$, $\tau = t\Gamma_2 / \pi R^2$.

For the pair of vortices similar to the Icelandic Low and Azores High in the Northern Hemisphere over the Atlantic Ocean (with z ~ 55º and y ~ 25º according to (Mokhov & Khon, 2005)) the stability condition (4.3) for $R_0$=0 is reduced to the following restrictions for the ratio between intensities of the vortices (for $\gamma_1 = -|\gamma_1| < 0$): $1.35 < |\gamma_1| < 5.24$. According to equation (4.1) the values used for $y$ and $z$ the value of $\gamma_0$ must have the form $\gamma_0 = 0.3(1.94 - |\gamma_1|)$ (Mokhov et al., 2013).

The expression for the period of the small oscillations for the vortices about equilibrium established according to (4.1), (4.2) with $\gamma_1 \neq -1$, $\sin z > \sin y$, and condition (4.3) has a form

$$T = \frac{2\pi \sin^2(z-y) \sin z \sin y}{\sqrt{-D}}. \qquad (4.4)$$

According to (4.4) the value $T$ may take arbitrary large values and describe respective long-period small oscillations near equilibrium if $|D| \to 0$. In particular, it may represent the relationship between vortices similar to the Azores High and Icelandic Low (for values $|\gamma_1| \to 1.35$ or $|\gamma_1| \to 5.24$).

The noted important role of polar vortex pair in the case $\gamma_0 \neq 0$, however, is not as principally significant as in the case for $\gamma_1 = -1$ with $\sin z \neq \sin y$ (i.e. for $z \neq \pi - y$) when the stationary mode (4.1), (4.2) may be realized only in the presence of a finite value parameter $\gamma_0$, characterizing the relative intensity of the polar vortex pair. For the constant in time value $\Gamma_0$ the stationary vortex is always unstable with respect to small disturbances.

Let us consider now in the general case a problem of stability of the constant velocity zonal vortex pair motion, the solution for the particular case of which is given above (see (3.13) – (3.18)). The condition of that stationary mode realization with N=2 for the system (3.4) is given by the following equalities

$$g_1 = \frac{\sin(\theta_{20} - \theta_{10})}{\sin \theta_{20}} \left[ (1 + \omega_0) \sin^2 \theta_{20} - g_0 \right],$$
$$g_2 = -\frac{\sin(\theta_{20} - \theta_{10})}{\sin \theta_{10}} \left[ (1 + \omega_0) \sin^2 \theta_{10} - g_0 \right]$$
(4.5)

The condition (4.5) in the particular case (3.13) corresponds to the condition (3.14) and equality $g_2 = -g_1$. For the zero zonal transfer angular velocity, when $\omega_0 = 0$, the condition (4.5) coincides with the conditions (4.1), (4.2).

It is clear that a stability condition of that stationary mode of the zonal vortices transfer is also inequality (4.3), in which $\gamma_1 \equiv \frac{g_1}{g_2}$ with $g_1$ and $g_2$ from (4.5). The condition of zonal motion the west direction not necessarily shall correspond to the requirement of negativity of $g_1$ in (4.5) (that may be satisfied also in (4.5) only when $\omega_0 = -|\omega_0| < 0; |\omega_0| > 1; g_0 > 0$). For the positive $g_1 > 0$ in (4.5), for the particular case $\theta_{10} = 35^0, \theta_{20} = 45^0$, from (4.3) and (4.5), we get the following stability condition of zonal pair of APV transfer in the west direction

$$0 > \omega_0 > -(1 - 5.89 g_0);$$
$$g_0 < 0.17(1 - |\omega_0|)$$
(4.6)

For example, the value $\omega_0 \approx -0.005$ defines respective restriction on the dimensionless value of the polar vortex intensity, $g_0 < 0.169$, when the stable pair of APV transfer in the west direction is possible.

On the base of the presented solution of the dynamic system (3.4), there is an opportunity for modelling not only stationary modes of ACA over the oceans type but also stationary modes of zonal vortices transfer typical for block events. However, a problem emerges related with the fact that for the example in (4.6), the value of $g_1 > 0$. It corresponds to cyclonic circulation of the vortex placed to the north in the APV vortex pair under consideration moving to the west. If we require in addition to the stability condition also the request for $g_1 < 0$, then instead of the condition (4.6), for the same latitude of the vortex pair position, we get

$$-0.17(|\omega_0|-1) < g_0 < 0.17(|\omega_0|-1) . \tag{4.7}$$

In (4.7), it is assumed that the motion in the west direction is realized with sufficiently large above-threshold angular velocity with $|\omega_0| > 1$. This condition is related with the exact accounting for the sphere rotation impact on the point vortices dynamics on the sphere.

In the case when it is possible neglecting of the sphere rotation and presence of the polar vortices, it follows from (4.5) that

$$\begin{aligned} g_1 &= \omega_0 \sin\theta_{20} \sin(\theta_{20} - \theta_{10}), \\ g_2 &= -\omega_0 \sin\theta_{10} \sin(\theta_{20} - \theta_{10}) \end{aligned} \tag{4.8}$$

From (4.8) it follows that only under condition of moving of a vortex pair in the west direction (for $\omega_0 < 0$ in (4.8)), is sufficient for the vortex number 1, positioned to the north (i.e. for $\theta_{10} < \theta_{20}$) with respect to the vortex number 2, to have the anti-cyclonic circulation with $g_1 < 0$. It corresponds to the condition of realization of a block of split flow type. As it was noted above, for non-zero polar vortices intensity, the stationary mode of zonal vortex pair transfer described in (4.8), is stable. The results obtained can be used for analysis of the block modes. According to (4.8), for sufficiently small by absolute value negative values, $\omega_0 < 0$, the negative following value, $G = g_1 \cos\theta_{10} + g_{20} \cos\theta_{20} = \omega_0 \sin^2(\theta_{20} - \theta_{10})$, also may be small by absolute value, that complies to the real blocks data. That conclusion is opposite to the conclusion in (DiBatista & Polvani 1998) on instability of the stationary mode of zonal motion of a vortex pair under small absolute values $|G| < 0.1$.

## 5. Discussion and conclusions

In the framework developed in the present paper, the approach for vortex dynamic analysis on a uniform rotating sphere, the non-uniformity of the underlying surface, the non-adiabatic processes, and other factors typical in a real system of atmospheric vortices were not taken into account. Nevertheless, it is possible to estimate significance of the contribution for the dynamic component of vortex interactions into the evolution of large-scale cyclonic and anticyclonic ACAs that play an important role in the Earth climate system (Mokhov & Petukhov 2000; Mokhov & Khon 2005; Khon & Mokhov 2006). A trial describing the ACA dynamics with the use of point vortices was undertaken, in particular, in (Bogomolov 1979), but without estimation for significance of the contribution of different factors from observations. One of the factors not taken into account in the proposed theoretical analysis above, but which should have an effect on the relative dynamics and stability of an ACA vortex system is the mean temperature difference between the oceans and continents. It is possible to analyze the mutual positions of ACA in the Northern Hemisphere taking into account variations of surface air temperature over the land and ocean (Mokhov et al. 2012).

An analysis was conducted in (Mokhov et al., 2012) for the variability in the mean distance between ACAs by longitude, $\Delta\lambda$ for the Atlantic (Icelandic Low and Azores High) and Pacific (Aleutian Low and Hawaiian High) ACAs as abnormality characteristics of the position and instability of the vortex pairs. In particular, reanalysis data were used for the detection of ACA characteristics similar to Mokhov & Khon (2005). The abnormality (instability) of mutual ACA positioning during the particular season was characterized in (Mokhov et al., 2012) by deviation from the long-term mean against the standard deviation. Also the degree of abnormality (instability) in the temperature difference between land and ocean $\Delta T$ in the Northern Hemisphere was estimated with respect to the mean conditions derived from the observations for winters from CRU data (http://www.uea.ac.uk/cru/data). The obtained estimates show the comparability of the dynamic and thermal factors in the formation of stability modes or instability of the ACA mutual positioning on the sphere (see also Appendix B). Figures 2-4 in Appendix B show stability regions depending on the positions of anticyclonic and cyclonic vortices on the sphere. The horizontal axis on Figs. 2-4 corresponds to the co-latitude $\theta$ for the anticyclonic ACA vortex center, and the vertical axis for that of the cyclonic ACA. Crosses on figures characterize mean values of $\theta$ for the respective vortices of the ACA pair (by the cross position) and their standard deviations (by the cross size in the respective direction). Dark region shows stability obtained according to the condition (4.3) for the particular parameter $\gamma_1$ values under consideration. The stationary mode of the vortex pair on Fig. 3 (similar to Icelandic-Azores ACA pair in 1988) falls in the stability region, and on Fig. 4 (similar to Icelandic-Azores ACA pair in 1964) does not fall into it. Shadowed regions on Figs. 2-4 correspond to the conditions of realization of exactly anti-cyclonic circulation of the polar vortex in the Northern hemisphere. It is clear that the value and sign of circulation of the polar vortices may be important for realization of a stable mode (see also (4.5) since defining stability parameter, $\gamma_1 = \dfrac{g_1}{g_2}$, explicitly depends on the value and sign of the polar vortex intensity, $g_0$).

Considered up to now only independently vortex regular waves and singular vortex dynamics on sphere are both the solutions of the absolute vorticity conservation equation (2.1) actually characterize the same hydrodynamic object which can be described by Legendre functions (of the first and second kind). This object is characterized by the some special dualism when point vortices correspond to the presence in some locations on the sphere of a singularity for these functions, but which are regular for the rest of the sphere. Here we considered the dynamics of such singularities (see also (Mokhov et al. 2012, 2013)) on a rotating sphere, at the first time exactly accounting for the sphere rotation impact in the obtained exact weak solution of AVCE. It allowed getting stationary modes corresponding o the stable stationary cyclonic-anticyclonic structures of ACA over oceans type.

As a result, it is possible using $N$-APV axes positions to model any number $N$-ACAs (in a form equivalent to the Barrett planetary vortices which also have static axes in the absolute coordinate system – see (2.6) for $\alpha = 0$). This model was realized in the present paper (with more details given for the case of $N = 2$). An exact accounting for the sphere rotation gives an opportunity to analyze block events on the base of the obtained stable modes of stationary zonal APV pairs motion.

The authors would like to thank Anthony R. Lupo for helpful comments. This research was supported by the Russian Science Foundation (grant 14-17-00806).

**AppendixA. Derivation of system (3.4) and its generalization**

A.1. Derivation of system (3.4)

It may be shown that the form of (3.1), (3.2) corresponds to an exact weak solution of equation (2.1) under condition that $\theta_i(t)$ and $\varphi_i(t)$ in (3.1), (3.2) satisfy a 2*N*-dimensional system of ordinary differential equations (3.4). Substitution of (3.1) in (2.1), multiplication of the differentiated result by an arbitrary finite function $\Phi(\theta,\varphi)$ and integration over the entire surface of the sphere results in the following:

$$\frac{1}{2}\int_0^{2\pi}d\varphi\int_0^{\pi}d\theta\sin\theta\Phi(\theta,\varphi)[\frac{\dot{\Gamma}_0}{R^2}(\delta(\theta)-\delta(\theta-\pi))+$$

$$\frac{1}{R^2}\sum_{i=1}^{N}[(\frac{\dot{\Gamma}_i}{\sin\theta_i}-\frac{\Gamma_i\cos\theta_i\dot{\theta}_i}{\sin^2\theta_i})(\delta(\theta-\theta_i)\delta(\varphi-\varphi_i)-\delta(\theta-\pi+\theta_i)\delta(\varphi-\varphi_i-\pi))+$$

$$\frac{\Gamma_i(-\dot{\theta}_i)}{\sin\theta_i}(\delta(\varphi-\varphi_i)\frac{\partial}{\partial\theta}\delta(\theta-\theta_i)+\delta(\varphi-\varphi_i-\pi)\frac{\partial}{\partial\theta}\delta(\theta-\pi+\theta_i))+$$

$$\frac{\Gamma_i(-\dot{\varphi}_i)}{\sin\theta_i}(\delta(\theta-\theta_i)\frac{\partial}{\partial\varphi}\delta(\varphi-\varphi_i)-\delta(\theta-\pi+\theta_i)\frac{\partial}{\partial\varphi}\delta(\varphi-\varphi_i-\pi))+$$

$$\frac{V_\theta\Gamma_i}{R\sin\theta_i}(\delta(\varphi-\varphi_i)\frac{\partial}{\partial\theta}\delta(\theta-\theta_i)-\delta(\varphi-\varphi_i-\pi)\frac{\partial}{\partial\theta}\delta(\theta-\pi+\theta_i))+$$

$$\frac{V_\varphi\Gamma_i}{R\sin^2\theta_i}(\delta(\theta-\theta_i)\frac{\partial}{\partial\varphi}\delta(\varphi-\varphi_i)-\delta(\theta-\pi+\theta_i)\frac{\partial}{\partial\varphi}\delta(\varphi-\varphi_i-\pi))]$$

(A.1)

=0.

After using integration by parts and taking into account the continuity equation $div\bar{V} = \frac{1}{R\sin\theta}(\frac{\partial V_\theta \sin\theta}{\partial \theta} + \frac{\partial V_\varphi}{\partial \varphi}) = 0$ in (A1), we obtain (taking into account zeroing of expressions near $\Phi(\theta,\varphi)$ and its same derivatives, and arbitrariness of the function $\Phi(\theta,\varphi)$) the following system is obtained:

$$\dot{\theta}_i = \frac{V_\theta}{R} = \frac{1}{R^2 \sin\theta_i}(\frac{\partial \psi}{\partial \varphi})_{\varphi=\varphi_i; \theta=\theta_i}$$

$$\dot{\varphi}_i = \frac{V_\varphi}{R\sin\theta_i} = -\frac{1}{R^2 \sin\theta_i}(\frac{\partial \psi}{\partial \theta})_{\varphi=\varphi_i; \theta=\theta_i}$$

(A.2)

From (3.2) and (A.2) we have the system (3.4) when $\dot{\Gamma}_i = 0$ (i.e. $\Gamma_i = const$, $i=1,2,..,N$) for any $\Gamma_0(t)$. Arbitrariness in the defining of the function $\Gamma_0(t)$ can be eliminated by considering in (3.5) and (3.8) the case when $\Gamma_0$ and $\Omega$ depend on time.

A.2. System (3.4) generalization

Let us get exact weak solution of the potential vorticity conservation equation $P = \frac{\omega}{H} = \frac{\Delta\psi + 2\Omega\cos\theta}{H}$ (see Section 2.1). We seek a solution of the equation (for stream function $\psi$ corresponding to the potential vortex $P$)

$$\frac{D}{Dt}(\frac{\Delta\psi + 2\Omega\cos\theta}{H}) = 0 \qquad (A.4)$$

In the form of a linear superposition of the stream functions $\psi = \psi_{0H} + \psi_H$ where stream function $\psi_{0H}$ is found as a solution of the following differential equation

$$\Delta \psi_{0H} + 2\Omega \cos\theta = P_0 H_0 h(\theta,\varphi,t)$$
$$h = \frac{H}{H_0}, H_0 = const, P_0 = const \quad (A.5)$$

In (A.5), a constant $P_0$ is introduced corresponding to a constant value of the potential vortex and is related with a representation of the full potential vortex value as follows: $P = P_0 + P_H, P_H = \frac{\Delta \psi_H}{H_0 h}$. Equation (A.4) implies

$$\frac{D}{Dt}\frac{\Delta \psi_H}{h} = 0, \quad (A.6)$$

To find an exact weak solution of (A.6) we use for the potential vortex, $P_H$, are presentation corresponding to (3.1) (in the case of $\Gamma_0 = 0$), when the stream function, $\psi_H$, is defined from the following equation

$$\Delta \psi_H = \sum_{n=1}^{N} \frac{\Gamma_{nh}}{R^2 \sin\theta_n}(\delta(\theta-\theta_n)\delta(\varphi-\varphi_n) - \delta(\theta-\pi+\theta_n)\delta(\varphi-\varphi_n-\pi)). \quad (A.7)$$

In the result, the form of the sum stream function, $\psi = \psi_{0H} + \psi_H$, exactly coincides with (3.2) if in (3.2) to replace constant intensities $\Gamma_n$ by the new constants $\Gamma_{nh} = \Gamma_n(t)h(\theta_n(t),\varphi_n(t),t) = const$, and the function $\psi_0$ replace by the stream function $\psi_{0H}$, that is a solution of the equation (A.5). When the dimensionless function, $h$, defining the fluid layer height is a given function of time and sphere surface coordinates, the exact weak solution of the equation (A.4) has a form of the system (3.4), in which, however, constant values are already not intensities of the point vortices, $\Gamma_n$, but the values of the following products: $\Gamma_n h(\theta_n(t),\varphi_n(t),t)$.

In the general case, the incompressible fluid layer height must be found from the following equation (of mass conservation):

$$\frac{\partial h}{\partial t} + \frac{V_\theta}{R}\frac{\partial h}{\partial \theta} + \frac{V_\varphi}{R\sin\theta}\frac{\partial h}{\partial \varphi} = 0 \ . \tag{A.8}$$

In (A.8), the velocity field, $V_\theta = \frac{1}{R\sin\theta}\frac{\partial \psi}{\partial \varphi}, V_\varphi = -\frac{1}{R}\frac{\partial \psi}{\partial \theta}$, is defined by the form of the stream function, $\psi = \psi_{0H} + \psi_H$, found from the equations (A.5) and (A.7). Generalization of the system of equations (3.4) and (A.2) is a system of the form (A.2) if to replace in (A.2) the stream function $\psi$ by $\psi = \psi_{0H} + \psi_H$, where $\psi_{0H}$ is found by solving equations (A.5), (A.8), and the function $\psi_H$ is defined from (A.7) and is as follows (see also 3.2)):

$$\psi_H = \sum_{n=1}^{N}\frac{\Gamma_{nh}}{\pi}Q_0(\cos u_n(\theta,\theta_n(t);\varphi,\varphi_n(t))); \tag{A.9}$$

where $\cos u_n = \cos\theta\cos\theta_n + \sin\theta\sin\theta_n\cos(\varphi-\varphi_n); Q_0(x) = \frac{1}{2}\ln(\frac{1+x}{1-x})$.

In particular, when in the equation (A.8), we can use approximation $\psi \approx \psi_{0H}$, equations (A.5) and (A.8) become independent from (A.9) and are reduced to the equation (2.1), allowing solution of the form (2.6). In the result, dynamics of the system of point vortices will be already defined taking into account function (2.6) instead of the stream function $\psi_0 = -\Omega R^2 \cos\theta$ used in the main text.

**Appendix B. APV stability regions**

Figures B.1-B.3 show stability regions (noted by the dark-gray color) depending on the co-latitude $\theta$ for the center of anticyclonic (horizontal axis) and cyclonic (vertical axis) vortices (in degrees) for different values of parameter $\gamma_1$, corresponding to the theoretical model of the stability criterion (4.3). Position (with standard deviations) of the vortex pair (similar to Icelandic-Aleutian ACA, in particular) is also presented. Shadowed region corresponds to the conditions of realization of anticyclonic circulation mode for the polar vortex in the northern hemisphere.

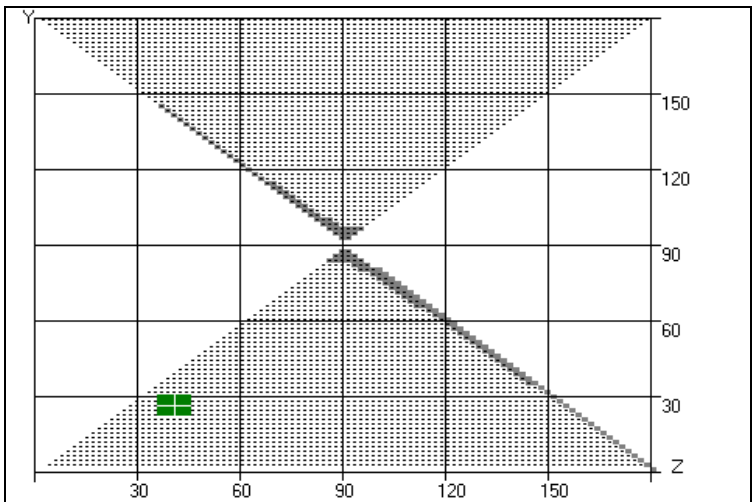

Fig. 2.

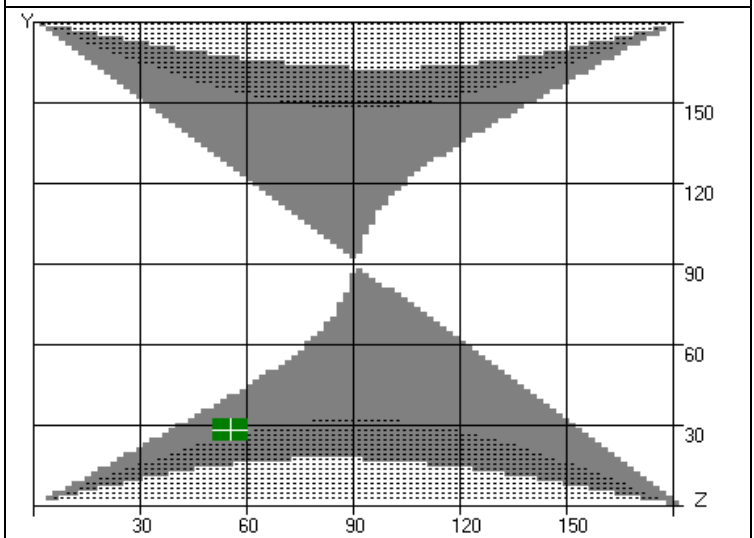

Fig. 3.

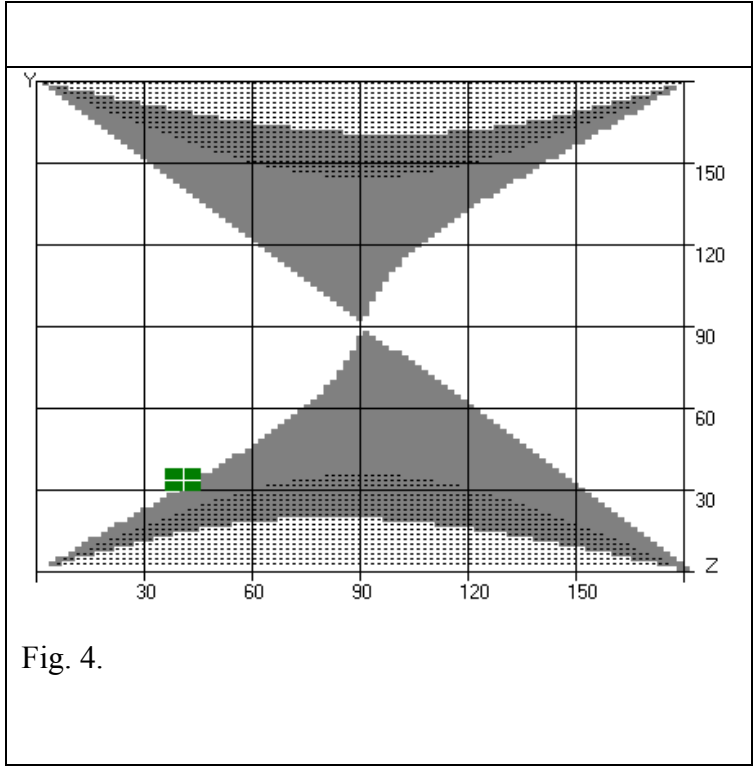

Fig. 4.

Fig. 2. Stability regions for $\gamma_1 = -1.000067$ with co-latitudinal position of vortex pair at $(27.5°, 42.5°)$.

Fig. 3. Stability regions for $\gamma_1 = -1.87$ with co-latitudinal position of vortex pair at $(30°, 57.5°)$ in the stability area.

Fig. 4. Stability regions for $\gamma_1 = -1.71$ with co-latitudinal position of vortex pair at $(35°, 42.5°)$ in the instability area.